\documentclass[fleqn,10pt]{wlscirep}
\title{Searchers adjust their eye movement dynamics to the target characteristics in natural scenes}
\usepackage{eurosym}

\author[1,*,+]{Lars O. M. Rothkegel}
\author[1,2,+]{Heiko H. Schütt} 
\author[1]{Hans A. Trukenbrod}
\author[2,3]{Felix A. Wichmann} 
\author[1]{Ralf Engbert} 

\affil[1]{Department of Psychology, University of Potsdam, Karl-Liebknechtstraße 24/25, 14476 Potsdam, Germany}
\affil[2]{Neural Information Processing Group, University of Tübingen, Sand 6,72076 Tübingen, Germany}
\affil[3]{Max Planck Institute for Intelligent Systems, Spemannstr. 34,72076 Tübingen, Germany}

\affil[*]{lrothkeg@uni-potsdam.de}

\affil[+]{these authors contributed equally to this work}

\begin{abstract}
When searching a target in a natural scene, both the target's visual properties and similarity to the background influence whether (and how fast) humans are able to find it. However, thus far it has been unclear whether searchers adjust the dynamics of their eye movements (e.g., fixation durations, saccade amplitudes) to the target they search for. In our experiment participants searched natural scenes for six artificial targets with different spatial frequency throughout eight consecutive sessions. High-spatial frequency targets led to smaller saccade amplitudes and shorter fixation durations than low-spatial frequency targets if target identity was known before the trial. If a saccade was programmed in the same direction as the previous saccade (saccadic momentum), fixation durations and successive saccade amplitudes were not influenced by target type. Visual saliency and empirical density at the endpoints of saccadic momentum saccades were comparatively low, indicating that these saccades were less selective. Our results demonstrate that searchers adjust their eye movement dynamics to the search target in a sensible fashion, since low-spatial frequencies are visible farther into the periphery than high-spatial frequencies. Additionally, the saccade direction specificity of our effects suggests a separation of saccades into a default scanning mechanism and a selective, target-dependent mechanism.
\end{abstract}
\begin{document}

\flushbottom
\maketitle

\thispagestyle{empty}

\section*{Introduction}

One of the most important everyday tasks of our visual system is to search for a specific target. Whether the task is to find a fruit amongst leaves, detect a dangerous animal or find relatives in a crowd of people, visual search has always been essential for survival. How our brain conducts this task so rapidly has been subject to a vast amount of research and, consequently, a number of comprehensive theories have been proposed \cite{treisman1980feature,wolfe1994guided,duncan1989visual}. However, most studies concerning visual search have been conducted on so-called search arrays, where targets and distractors are presented on a homogeneous background. Although the results from these studies are very useful for understanding the basic nature of visual search, many do not take eye movements into account, although eye movements play an important role in real world search behavior \cite{findlayactive,malcolm2009effects}. When searching on a complex background, saccades---fast ballistic eye movements---are executed about three to four times per second to increase the probability of finding a target. It has been shown in many studies that the search target strongly influence saccade target selection of eye movements when searching through natural scenes. Object-scene consistency \cite{loftus1978cognitive,henderson1999effects,cornelissen2017stuck}, scene context \cite{torralba2003modeling,neider2006scene} as well as low-level features \cite{hwang2009model} of the target influence where observers fixate. Thus, a top-down search template of the target appears to guide gaze during scene exploration \cite{wolfe1994guided,hwang2009model}. 
Correlations between the visual properties of target-related search templates and fixated image patches exist, but do not completely explain eye-movement behavior in visual search on complex backgrounds. Najemnik and Geisler \cite{najemnik2005optimal,najemnik2008eye} showed that human observers do not simply move their eyes to positions which maximally resemble the target (maximum a posteriori search) but rather apply a strategy which takes the visual degradation towards retinal periphery into account and samples as much information as possible with a minimal number of eye movements. Thus it seems useful for the visual system to adapt eye-movement strategies according to the target's visibility in the periphery. Target visibility depends on retinal eccentricity \cite{meinecke1989retinal} and its interaction with many factors such as spatial frequency \cite{pointer1989contrast} and contrast \cite{campbell1968application,robson1981probability}. 

To investigate whether the target features not only influence where participants look at (i.e. fixation locations) but also how they search (saccade amplitudes and fixation durations), we let participants search natural scenes for artificial targets with different low-level features. Although one might suspect that different targets lead to different saccade amplitudes and fixation durations, to our knowledge no one has yet provided empirical evidence to answer this question. It is rather important for models of eye movement control to find out whether, how fast, and how accurately human observers change search strategy according to the target they search for. To explicitly compare targets of different spatial frequency on various backgrounds, we used artificial targets instead of real-world objects in this study. We used scenes instead of plain backgrounds because i) we are interested in real-world search behavior and not search on highly controlled arrays and ii) to gain knowledge to improve dynamical models of saccade generation in natural scenes \cite{engbert2015spatial,schutt2017likelihood}. 

\begin{figure}
\unitlength1mm
\begin{picture}(100,100)
\put(40,0){\includegraphics[width=100mm]{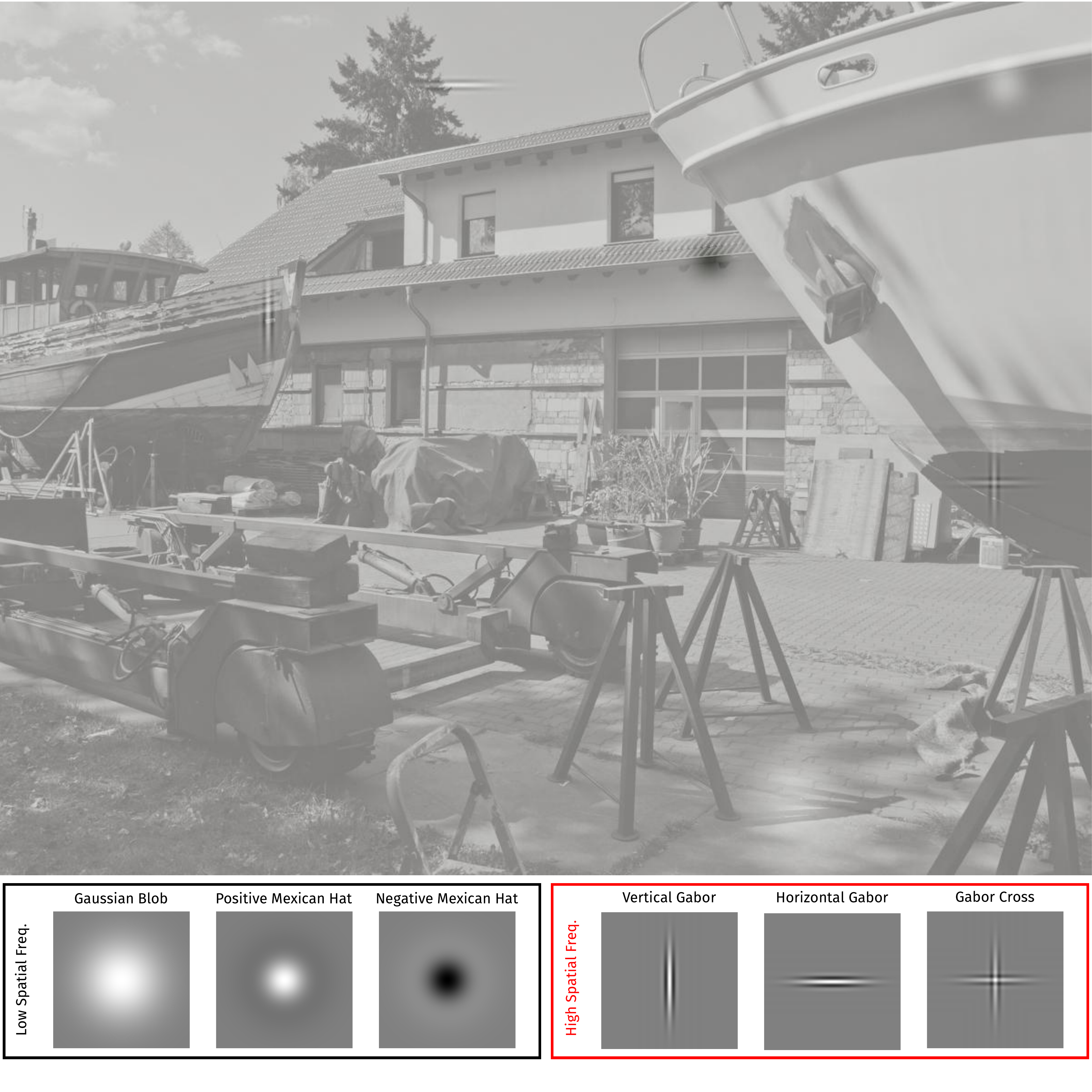}}
\end{picture}
\caption{\label{FigTargets}
Illustration of the task. Subjects were asked to search for one specific target for a block of 25 trials each overlayed over natural scenes like this one. In this image all 6 targets are hidden twice as large and at higher contrast than in the experiment to make them visible despite the small image. In the actual experiment only one target was hidden per image and the image was shown much larger. The bottom panels show the 6 targets we used. The frames around the targets mark which frequency category they belong to.}
\end{figure}

Observers searched in each of 8 consecutive sessions for 6 targets of varying spatial frequency content and, in the case of high-spatial frequency, orientation (vertical, horizontal, vertical+horizontal; see Fig.~\ref{FigTargets}). Each session contained one block per target. Each block consisted of one repetition of the same 25 images. Target type was specified in advance to each block to provide a search template. In one session (Session 7) targets were chosen randomly for each trial and target type was unknown prior to a trial. This was done to assure that differences in search behavior could be attributed to the search strategy and not the visual stimulus itself. 

If dynamical aspects of eye movements are indeed adapted to the search target in a useful way, saccade amplitudes should be higher for low-spatial frequency targets, since high-spatial frequencies can not be detected as far into the periphery as low-spatial frequencies \cite{pointer1989contrast}. High-spatial frequency targets are detected easier if they appear foveal \cite{schutt2017image}, thus it would make sense that fixation durations are shorter for high-spatial frequency targets. Additionally, because low-spatial frequency targets can be perceived from further away, and the size of the attentional window increases with longer fixation duration \cite{geisler1995separation}, it is also useful to prolong fixation durations for low-spatial frequency targets.

\section*{Results}

All figures show eye movement results for the different search targets. Bar plots (left side of figures 2--5) show overall results for the 6 targets and the results for all targets combined in Session 7, when target type was unknown. Line graphs in Figures 2--7 show comparisons between the three low-spatial frequency targets (Gaussian Blob and positive/negative Mexican hat, black line) and the three high-spatial frequency targets (vertical, horizontal bar and cross, red line; cf. Fig.~\ref{FigTargets}, bottom panels) throughout the course of the 8 experimental sessions. Error bars in the graphs represent the standard error of the mean. Significance signs always refer to differences between low and high-spatial frequency targets (* $p<.05$, ** $p<.01$, , *** $p<.01$). Solid lines below significance stars indicate significant differences for a range of neighboring data points.   

\subsection*{Task Performance}

\subsubsection*{Detection Rate}

\begin{figure}
\unitlength1mm
\begin{picture}(150,60)
\put(20,0){\includegraphics[width=6cm]{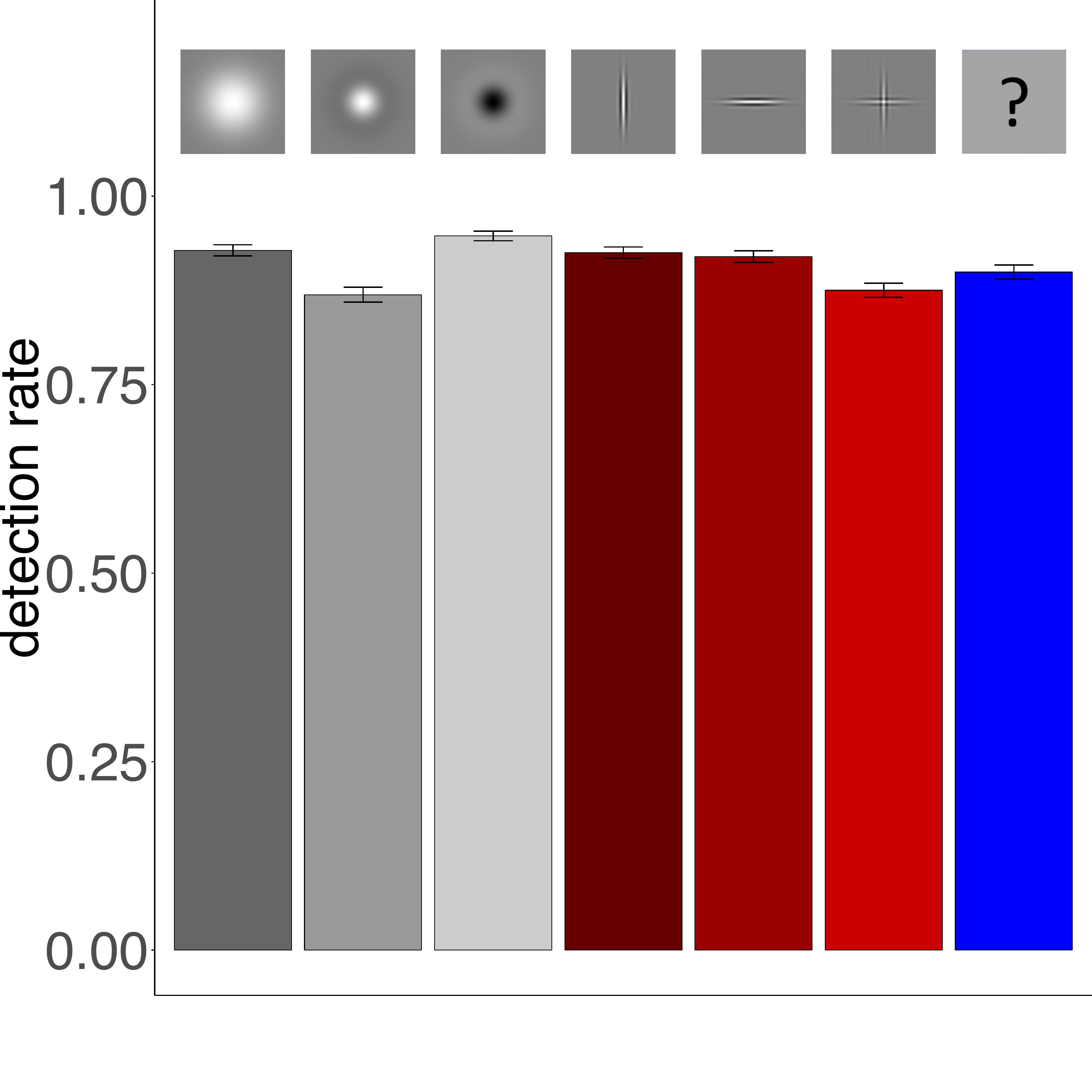}}
\put(85,0){\includegraphics[width=6cm]{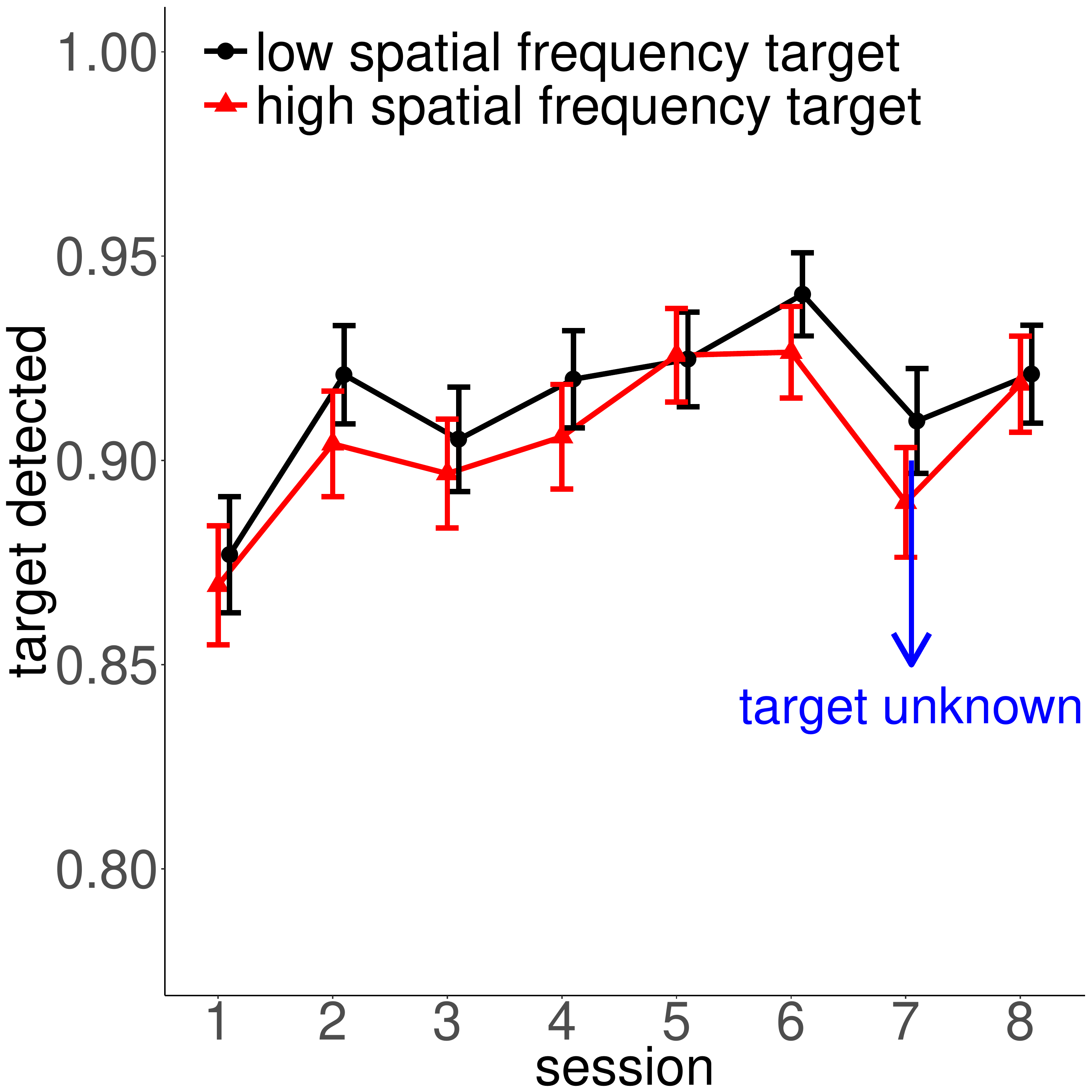}}
\put(20,60)A
\put(85,60)B
\end{picture}
\caption{\label{FigTaskPerformance}
A) Detection rate for the 6 targets. Red bars are high-spatial frequency targets and gray bars low-spatial frequency targets. The blue bar captures all trials where target type was unknown prior to the trial. B) Average detection rate of the three low and high-spatial frequency targets throughout the 8 experimental sessions. In Session 7 target type was unknown prior to a trial.}
\end{figure}

Participants showed similar detection rates (Hits/Misses) for the different targets throughout the whole experiment (Fig.~\ref{FigTaskPerformance} A).  The lowest detection rate  was observed for the positive Mexican hat and the high-spatial frequency cross (both 83\%) and the highest rate for the negative Mexican hat (92\%). The overall rate of false alarms was very low (3.44\% of target absent trials). Over the course of the experiment (Fig.~\ref{FigTaskPerformance} B), the detection rate for both, low and high-spatial frequency targets increased. No clear difference between the groups of high-spatial frequency and low-spatial frequency targets was observed. In Session 7, when target type was unknown prior to the trial, detection rates dropped for both target types but performance was still better than in the first experimental session.

\subsubsection*{Search Time}

\begin{figure}
\unitlength1mm
\begin{picture}(150,60)
\put(20,0){\includegraphics[width=6cm]{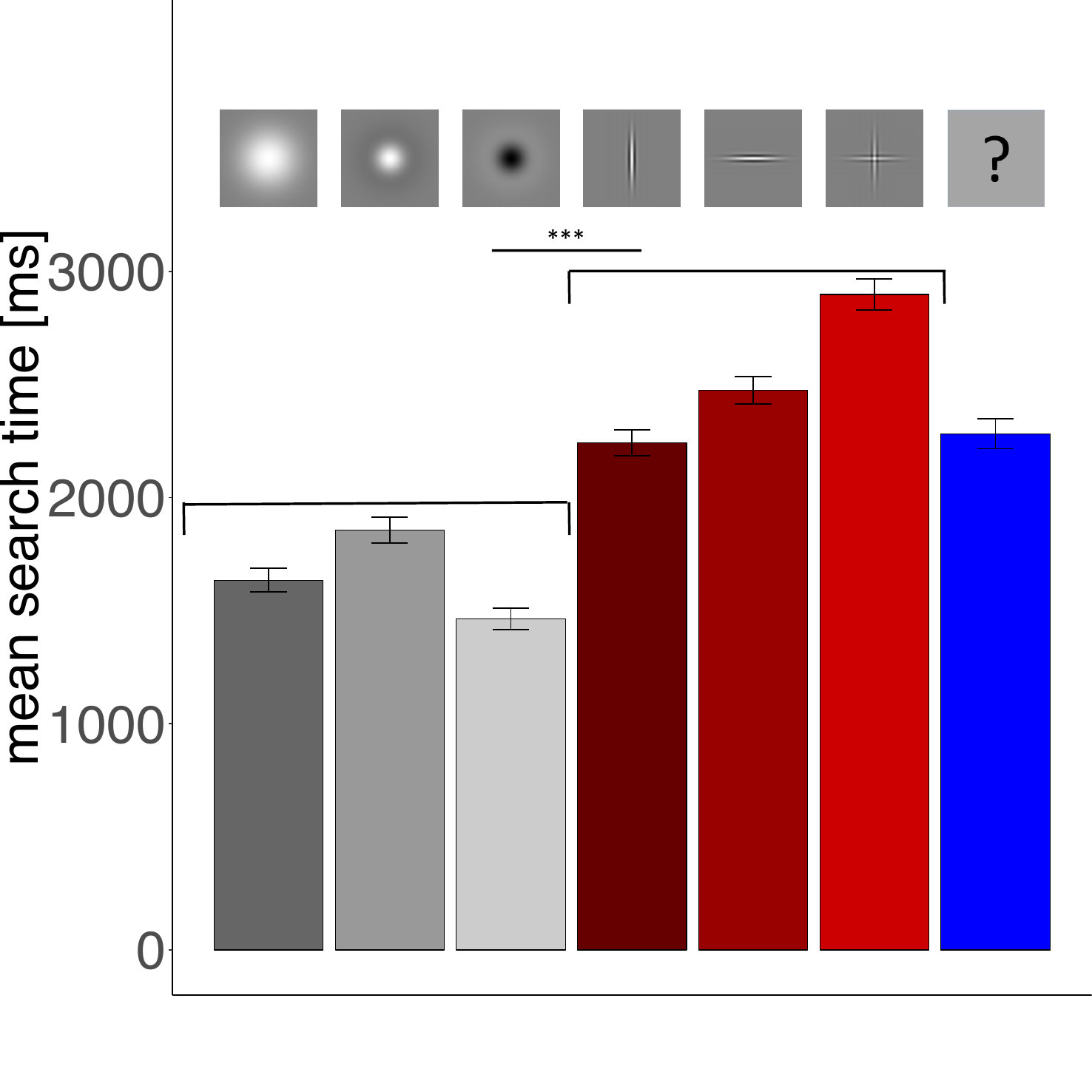}}
\put(85,0){\includegraphics[width=6cm]{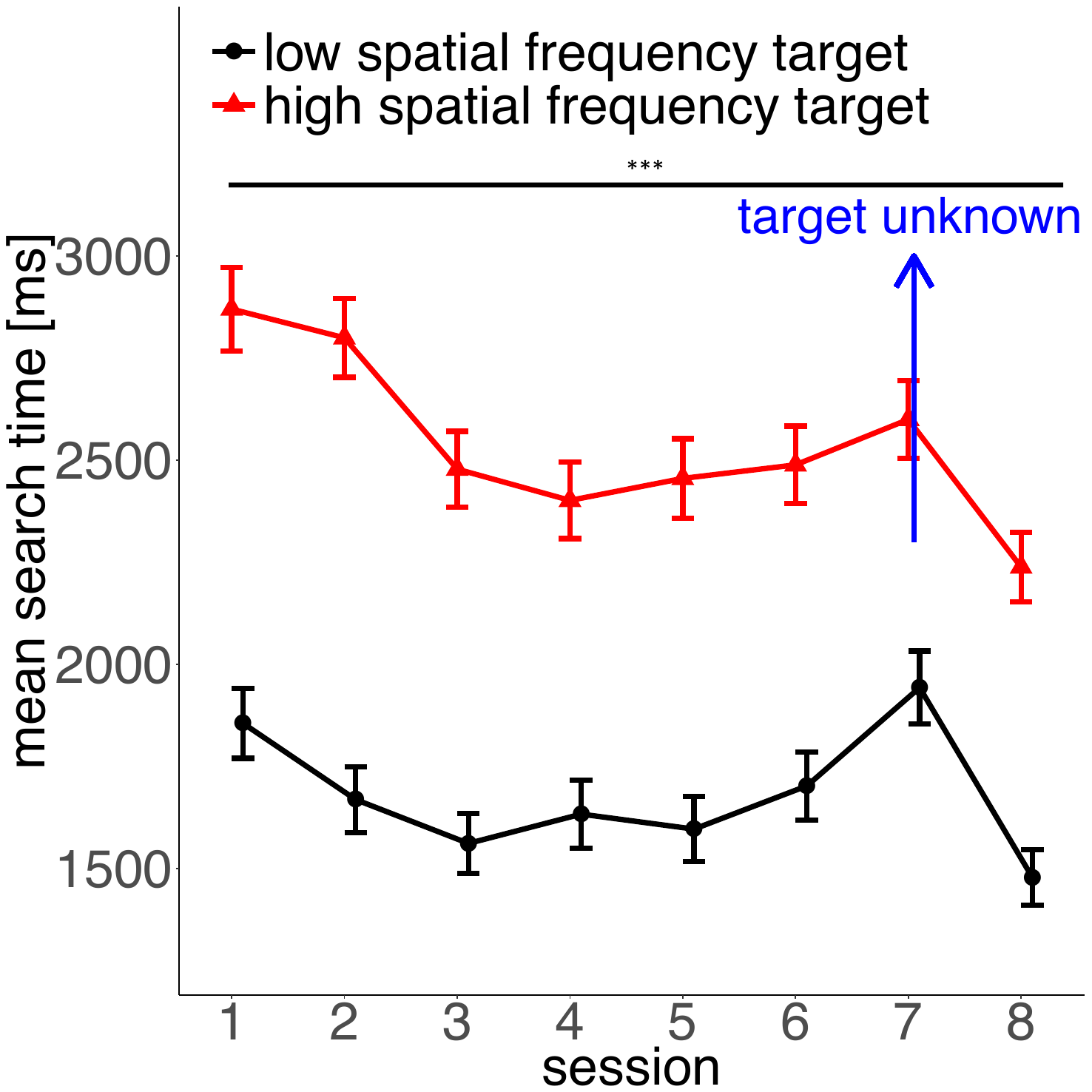}}
\put(20,60)A
\put(85,60)B
\end{picture}
\caption{\label{FigSearchTime}
A) Search times for the 6 targets. Red bars are high-spatial frequency targets and gray bars low-spatial frequency targets. The blue bar captures all trials where target type was unknown prior to the trial. B) Average search times for the three low and high-spatial frequency targets throughout the 8 experimental sessions. In Session 7 target type was unknown prior to a trial. In Session 7 target type was unknown prior to a trial.}
\end{figure}

The search time (Fig.~\ref{FigSearchTime} A) was more variable between the targets than the detection rate. Participants were faster at finding low-spatial frequency targets than high-spatial frequency targets. Participants were fastest at finding the negative Mexican hat and slowest at finding the high-spatial frequency cross. Analogous to the rise in detection performance, search time decreased over the 8 sessions. The first 3 sessions showed a clear training effect and afterwards a plateau was reached (Fig.~\ref{FigSearchTime} B). In Session 7 (target unknown) search times rose but high-spatial frequency targets were still detected faster than in the first session, indicating that search training compensated for loss of guidance in this case, which was also visible in detection performance. 

\subsection*{Scanpath Properties}

\subsubsection*{Saccade Amplitudes}

\begin{figure}[htb]
\unitlength1mm
\begin{picture}(150,60)
\put(20,0){\includegraphics[width=6cm]{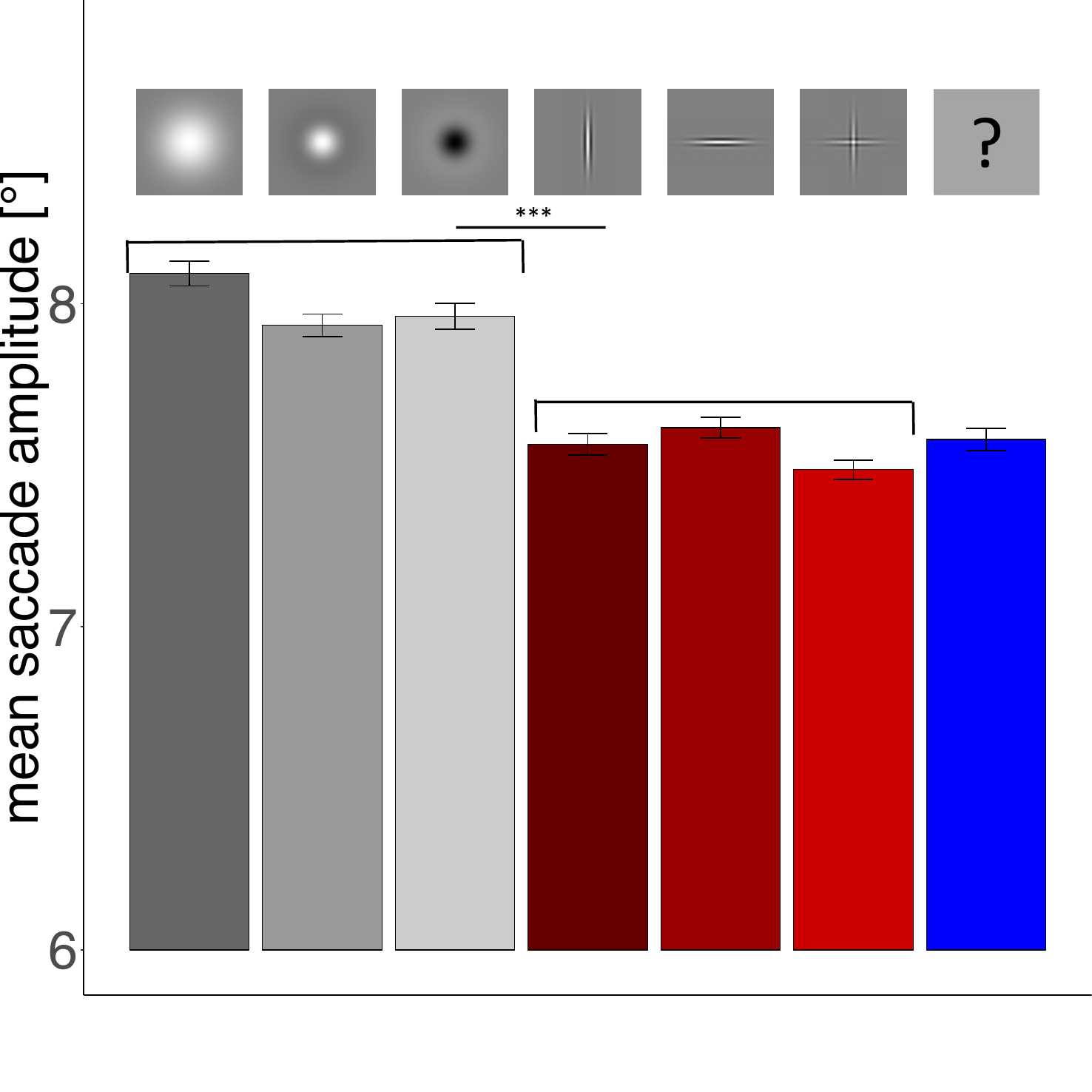}}
\put(85,0){\includegraphics[width=6cm]{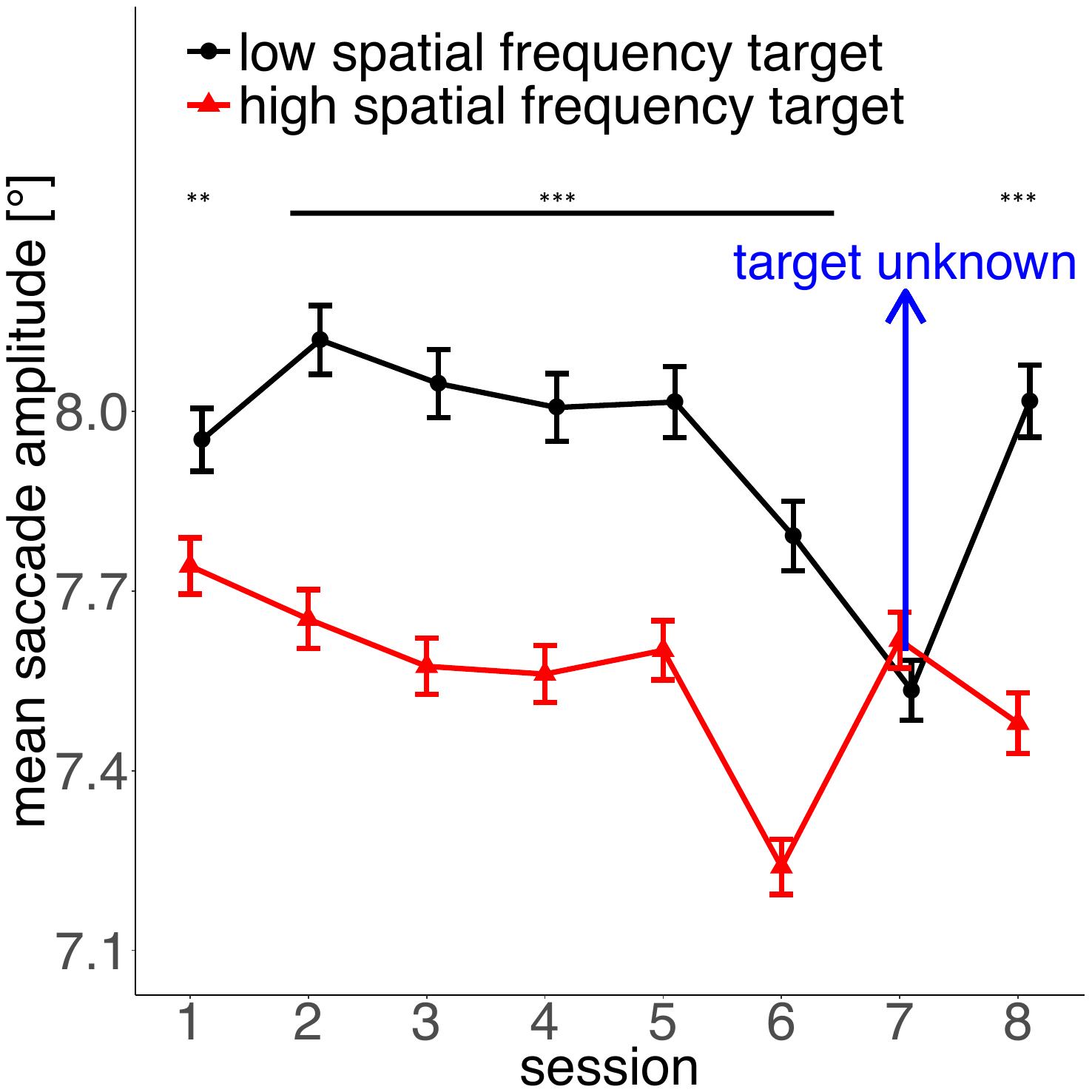}}
\put(20,60)A
\put(85,60)B
\end{picture}
\caption{\label{FigSaccAmp}
A) Mean saccade amplitude for the 6 different targets. Red bars are high-spatial frequency targets and gray bars low-spatial frequency targets. The blue bar captures all trials where target type was unknown prior to the trial. B) Average saccade amplitude of the three low and high-spatial frequency targets throughout the 8 experimental sessions. In Session 7 target type was unknown prior to a trial.}
\end{figure}

Analyses of the saccade amplitudes throughout our experimental sessions (Fig.~\ref{FigSaccAmp} A \& B) showed three clear results: (i) amplitudes were longer for low than for high-spatial frequency targets, (ii) this difference was established in the first session, persisted throughout all other sessions and (iii) vanished when target type was unknown prior to a trial. The Gaussian blob produces even larger amplitudes than the Mexican hats (Fig.~\ref{FigSaccAmp} A). This is strategically useful, because the blob is visible in an even further periphery than the Mexican hats.

\subsubsection*{Fixation Duration}

\begin{figure}[htb]
\unitlength1mm
\begin{picture}(150,60)
\put(20,0){\includegraphics[width=6cm]{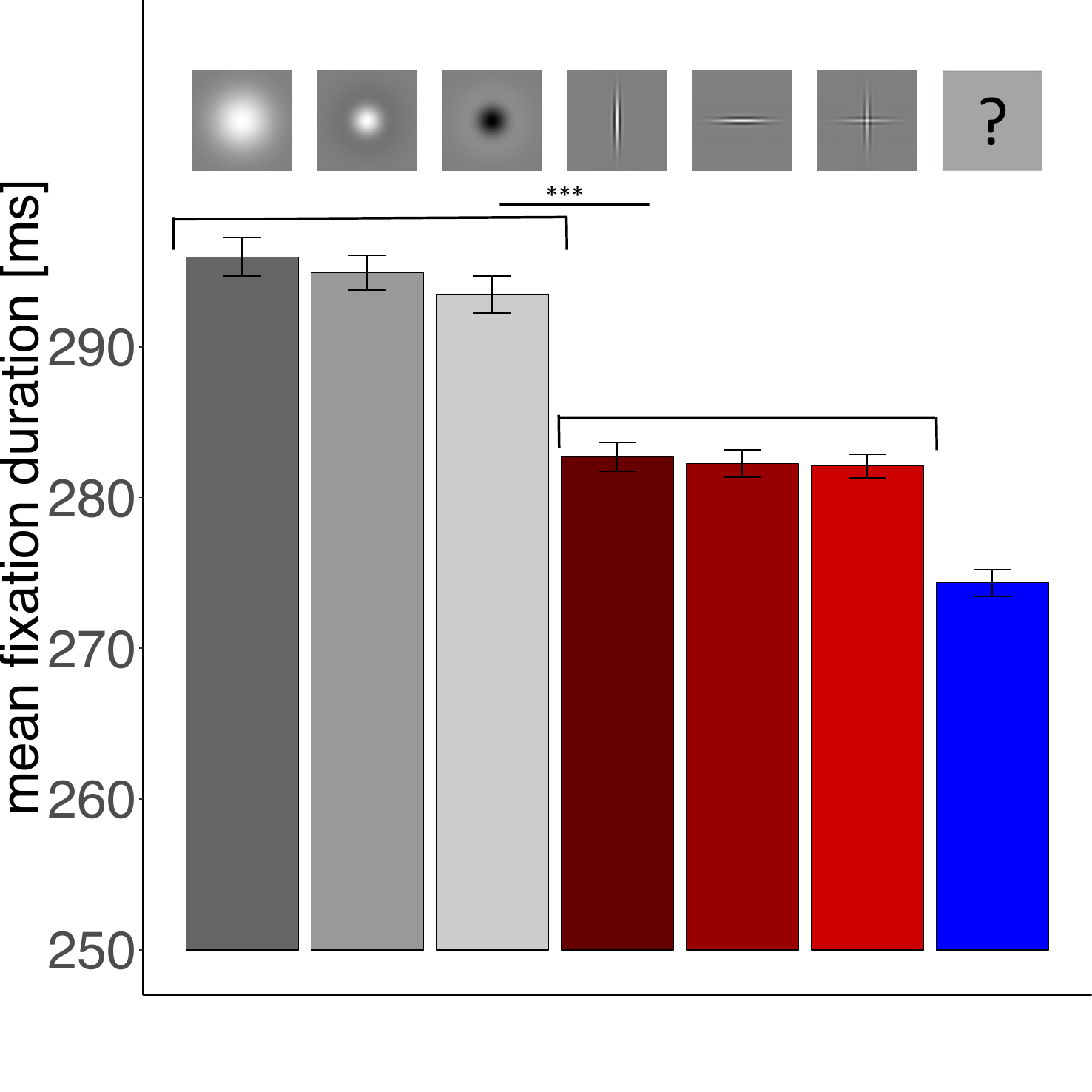}}
\put(85,0){\includegraphics[width=6cm]{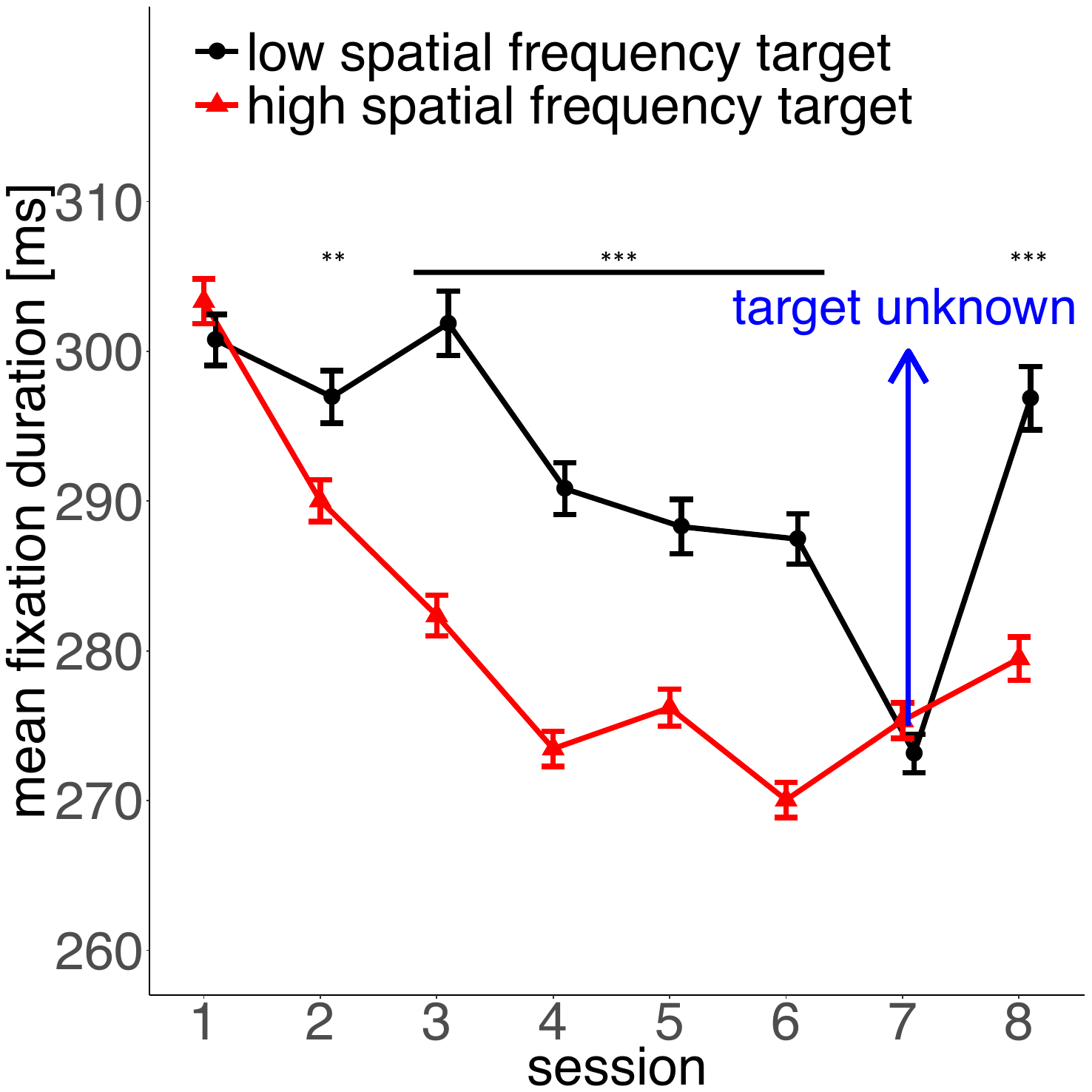}}
\put(20,60)A
\put(85,60)B
\end{picture}
\caption{\label{FigFixDur}
A) Mean fixation durations for the 6 different targets. Red bars are high-spatial frequency targets and gray bars low-spatial frequency targets. The blue bar captures all trials where target type was unknown prior to the trial. B) Average fixation duration of the three low and high-spatial frequency targets throughout the 8 experimental sessions. In Session 7 target type was unknown prior to a trial.}
\end{figure}

The pattern for fixation durations (Fig. \ref{FigFixDur}) was similar to the pattern of saccade amplitudes: (i) the three low-spatial frequency targets led to a search strategy with longer fixation durations, (ii) this difference in fixation durations needed one training session to be established, but afterwards persisted throughout the other sessions and (iii) vanished when target type was unknown prior to a trial. Interestingly, fixation durations are very short in the session when prior was unknown (Fig. \ref{FigFixDur} B). This might be because fixation durations decrease throughout the experiment and this session was the seventh of eight session but also indicates that participants chose the strategy of searching for high-spatial frequency targets, when they don't know what to look for. 

\subsubsection*{Time-course during a trial}
\begin{figure}
\unitlength1mm
\begin{picture}(150,60)
\put(20,0){\includegraphics[width=6cm]{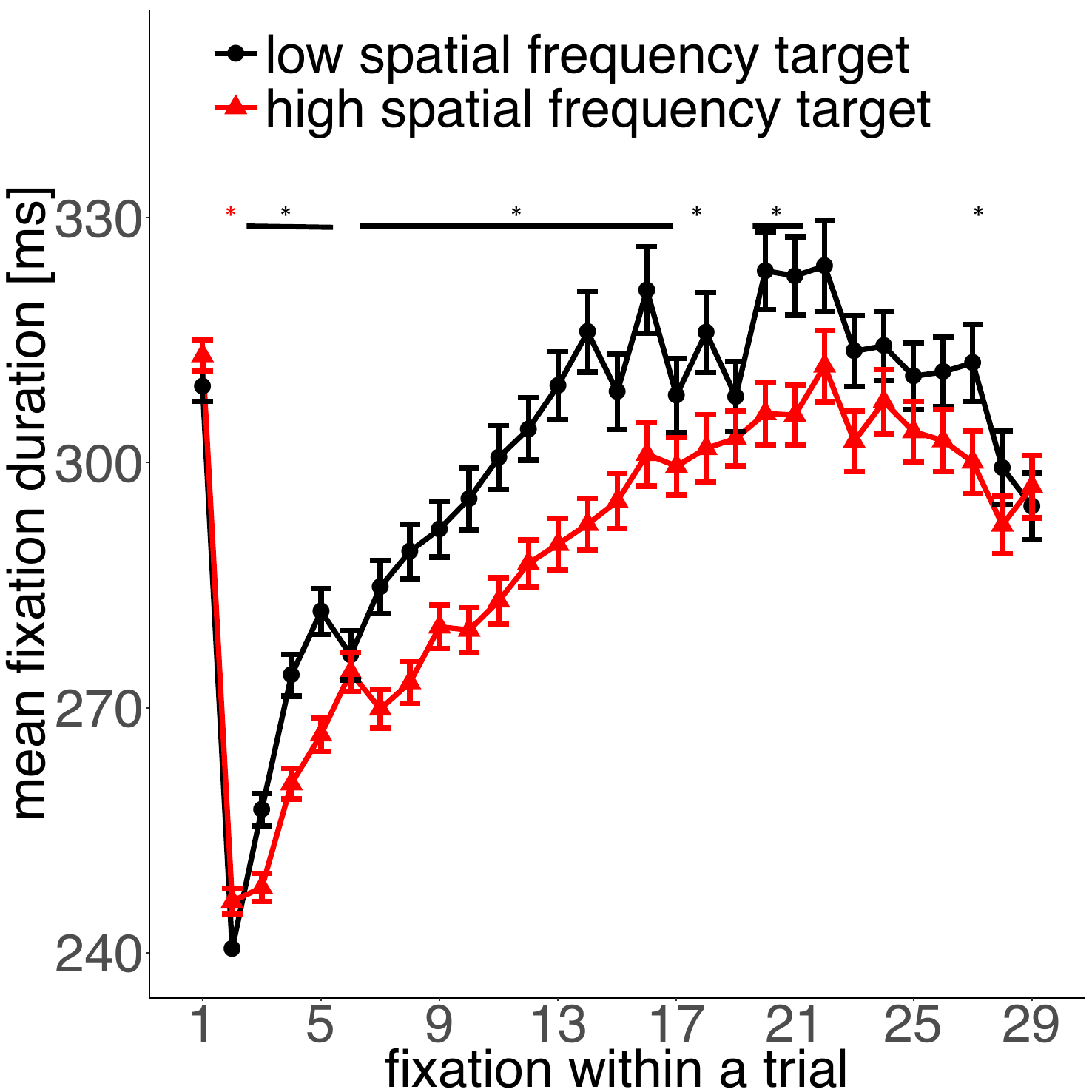}}
\put(85,0){\includegraphics[width=6cm]{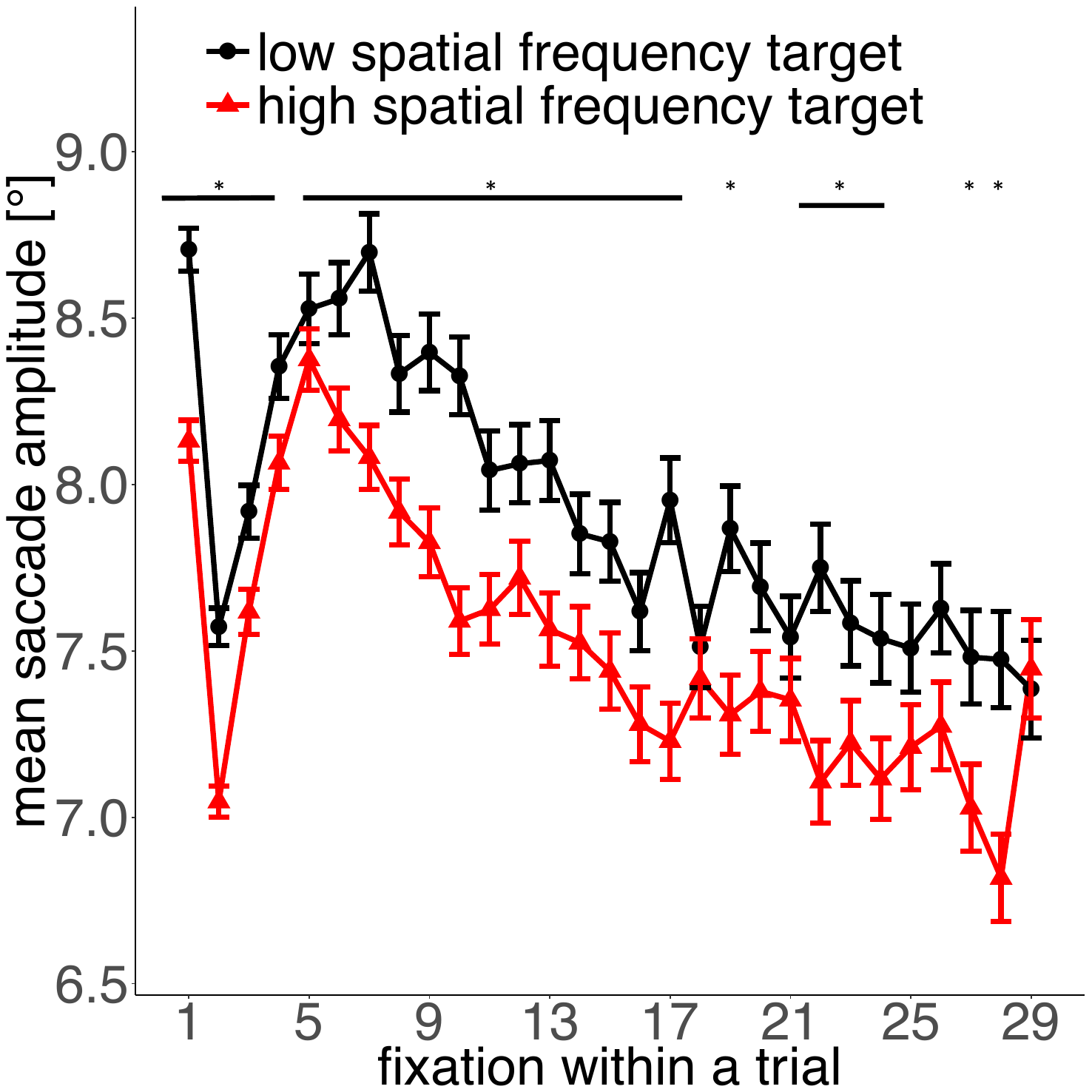}}
\put(20,60)A
\put(85,60)B
\end{picture}
\caption{\label{FigTrialEvolution}
Temporal evolution of A) mean saccade amplitude and B) fixation duration for the two targets types throughout a trial.}
\end{figure}

Throughout a trial, fixation durations increased and saccade lengths decreased on average (except for the first movement, which was influenced by the experimental design and the central fixation bias; Fig. \ref{FigTrialEvolution}). This behavior is known as the coarse-to-fine strategy of eye movements \cite{antes1974time,over2007coarse}. However, the effect of target spatial frequency occured already after the second saccade and then persisted throughout the whole trial. Thus, searchers followed a coarse-to-fine strategy, but this does not explain the difference between targets.

\subsubsection*{Change in saccadic direction}

\begin{figure}
\unitlength1mm
\begin{picture}(150,110)
\put(5,55){\includegraphics[width=5cm]{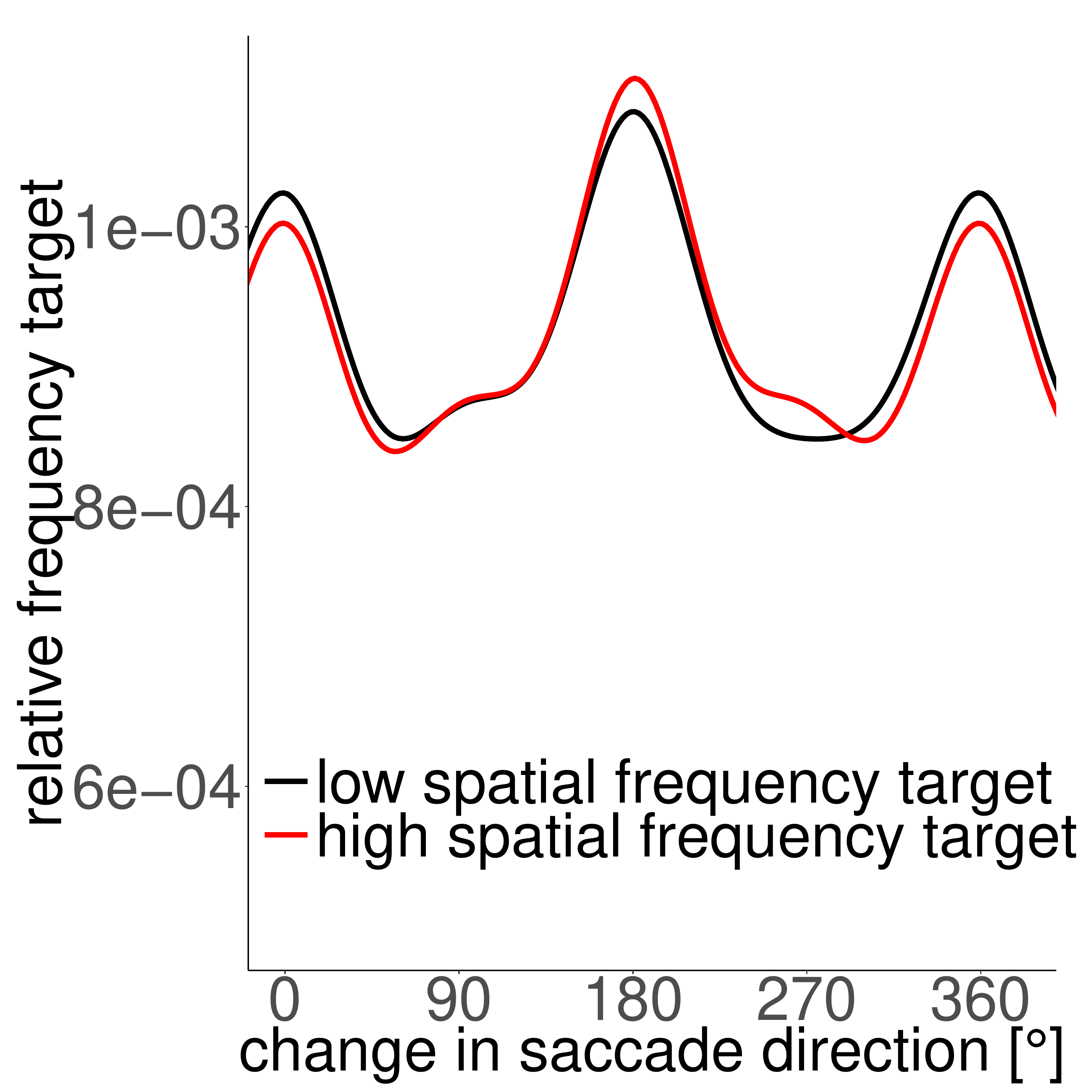}}
\put(60,55){\includegraphics[width=5cm]{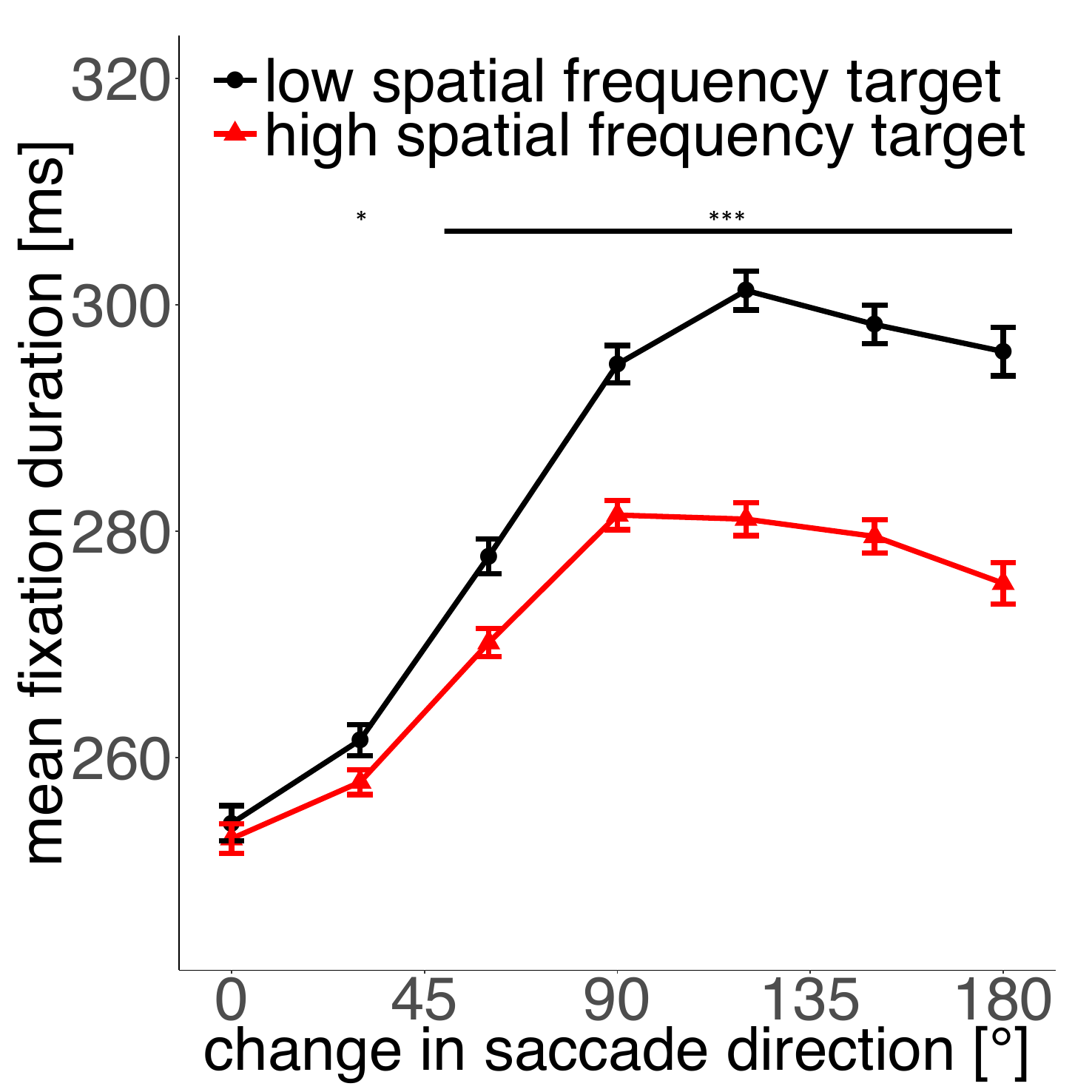}}
\put(115,55){\includegraphics[width=5cm]{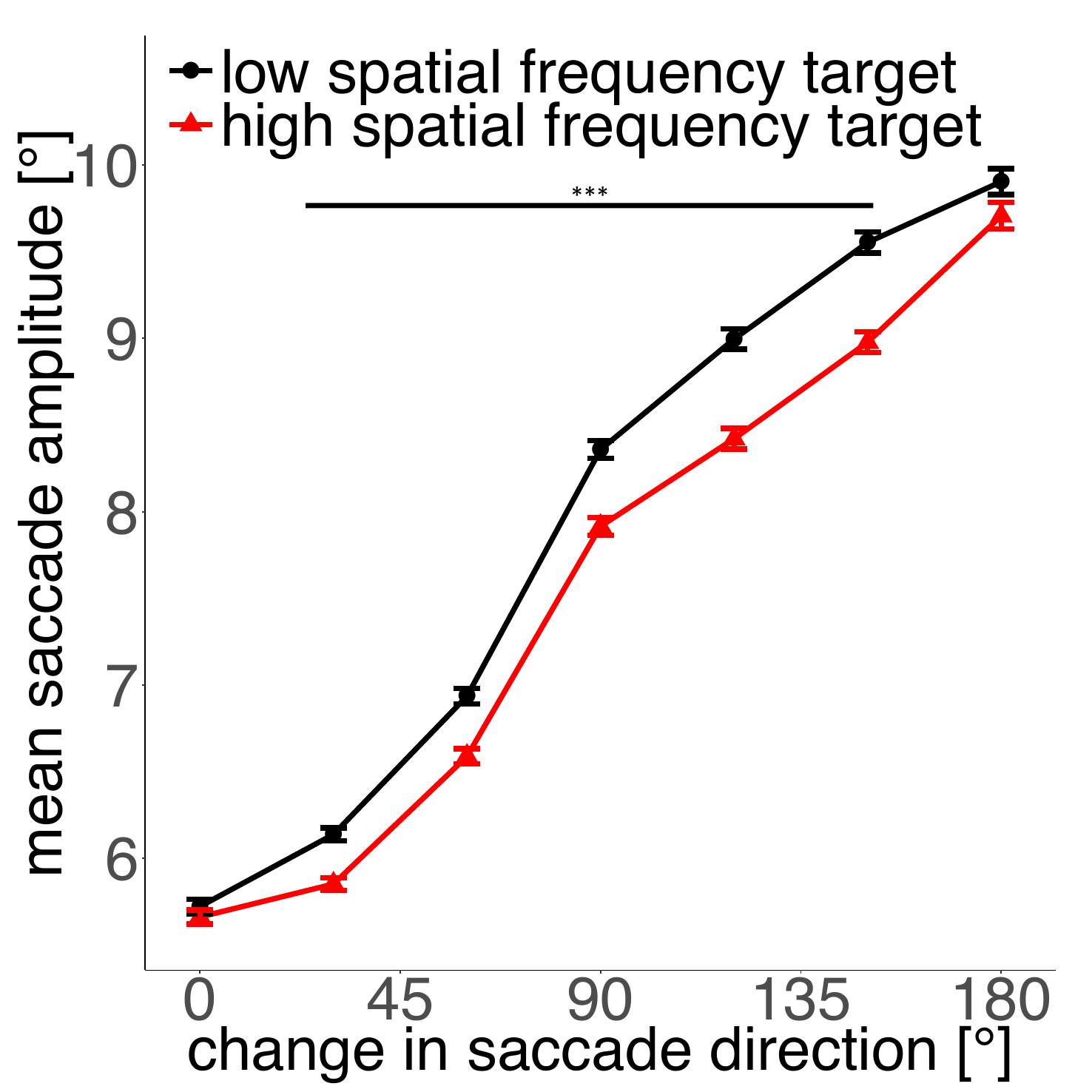}}
\put(35,0){\includegraphics[width=5cm]{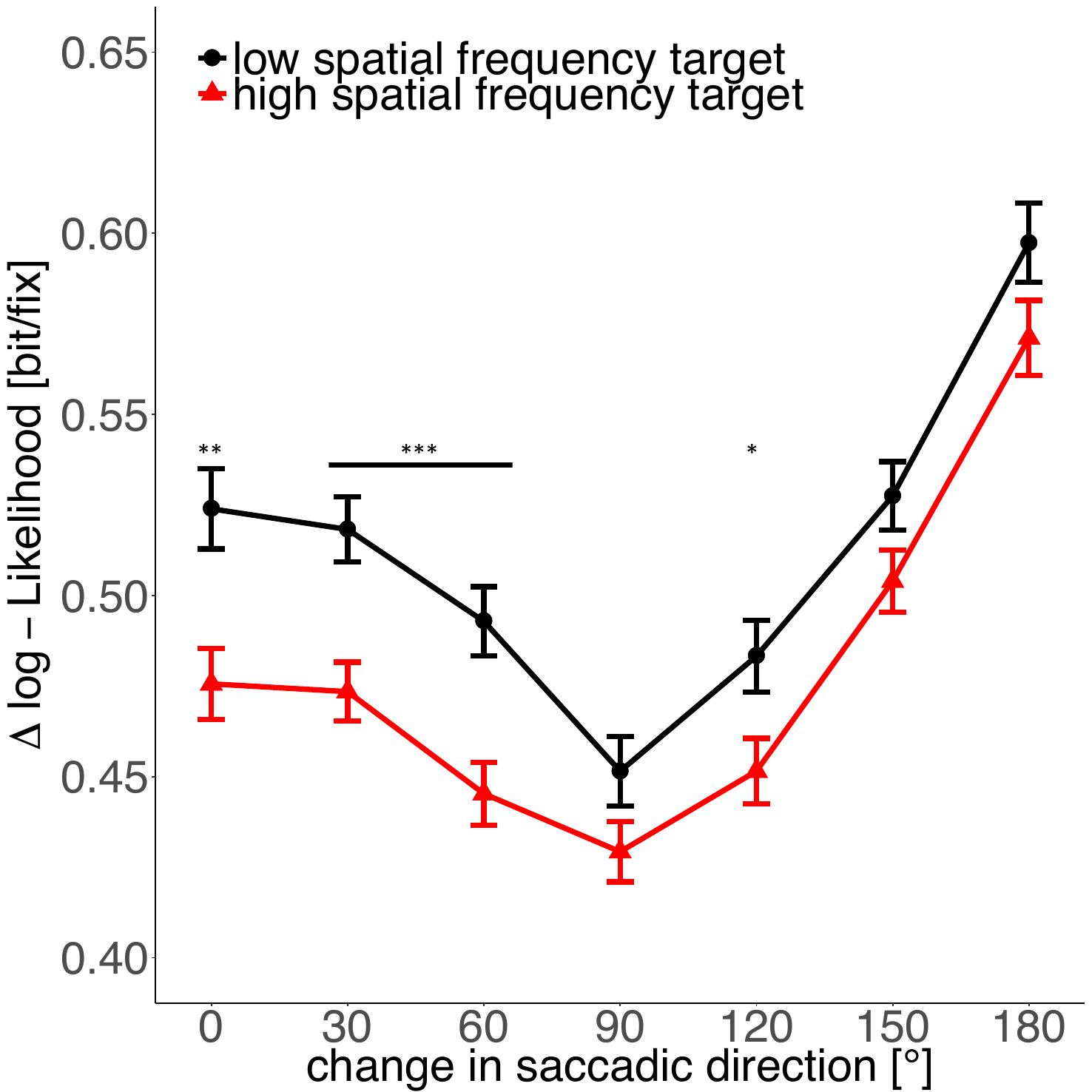}}
\put(85,0){\includegraphics[width=5cm]{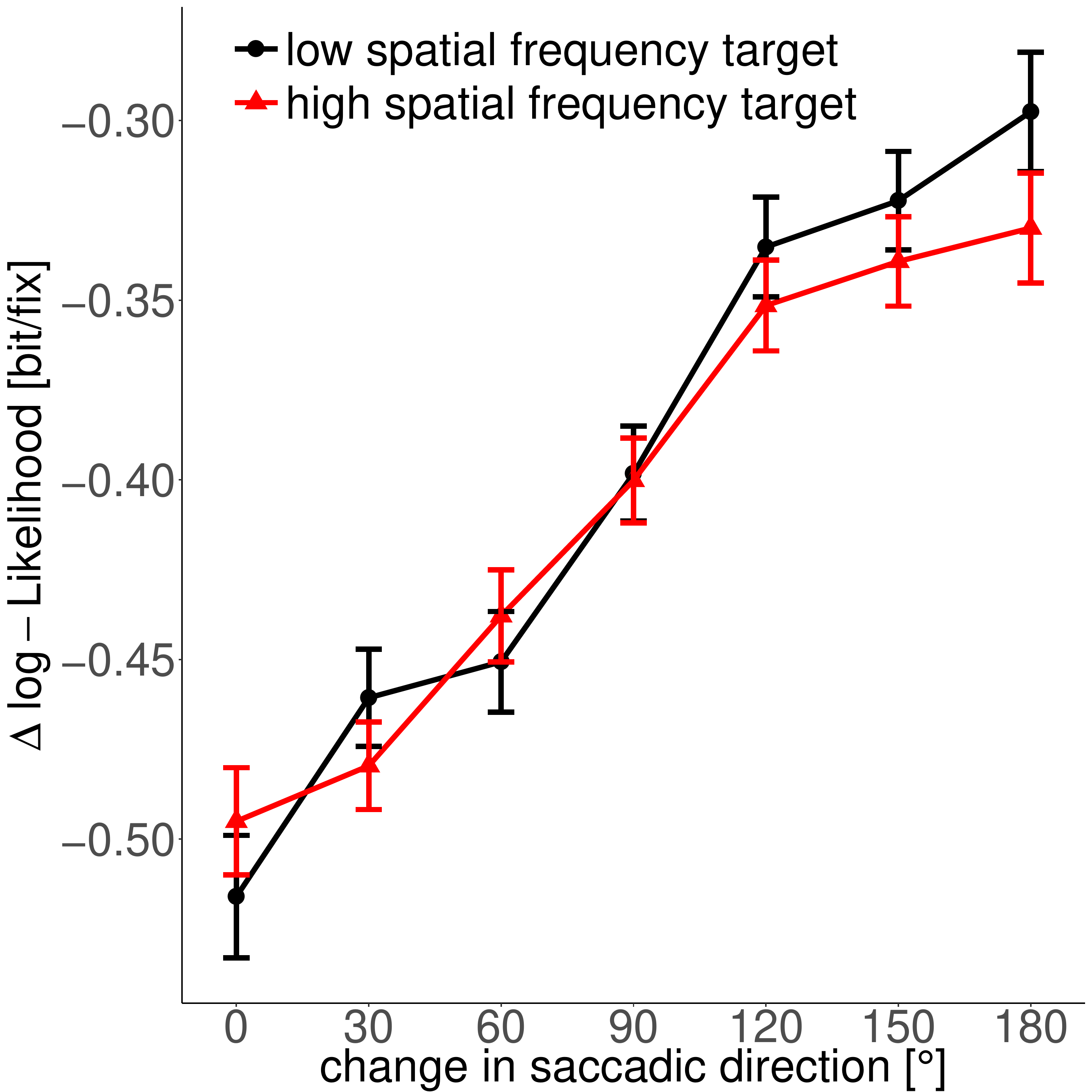}}
\put(5,105)A
\put(60,105)B
\put(115,105)C
\put(35,50)D
\put(85,50)E
\end{picture}
\caption{\label{FigAngle} A) Density distribution of change in saccadic direction. Influence of change in saccadic direction on B) fixation durations, C) successive saccadic amplitudes, D) empirical density of successive saccade target and E) DeepGaze saliency of successive saccade target.}
\end{figure}

Although we did not have a hypothesis about the interaction of saccade direction and visual search target, we included these analyses post-hoc, because the distribution of changes in saccade direction is a very systematic image-independent tendency \cite{tatler2009prominence,smith2009facilitation,rothkegel2016influence}. Because saccade amplitudes and fixation durations differed for the different targets, and the change in saccade direction has a large effect on saccade amplitude and fixation duration, this analysis promised interesting results \cite{tatler2008systematic,tatler2017latest}. Saccades which maintain direction (so called {\sl saccadic momentum} saccades) typically have small amplitudes and fixation durations between two saccades in the same direction are short. Backward saccades (or {\sl return saccades}) are usually large and fixation durations which preceed a change in direction are rather long. These effects were also found in our experiment (Fig.~\ref{FigAngle} B \& C). No apparent difference was found between the distributions of intersaccadic angles for low and high-spatial frequency targets (Fig.~\ref{FigAngle} A). 

Figure~\ref{FigAngle} C shows that saccades which maintain direction according to the previous saccade (0 degree change) are equally large for both target types. We interpret these saccades as less selective than saccades with a change in direction. This is even more evident if we look at fixation duration differences between the two target types (Fig.~\ref{FigAngle} B) in relation to the change in saccade direction. To further investigate this hypothesis, we compared the saccadic landing points for different changes in saccade direction in terms of empirical density and visual saliency. The empirical density maps were computed with the SpatStat package of the R language for statistical computing \cite{SpatStat,R} and for visual saliency we used the DeepGaze 2 model, which is highest ranking saliency model on the MIT saliency benchmark \cite{mit-saliency-benchmark,kummerer2014deep}.

Figures \ref{FigAngle}D and E show that empirical density and saliency, respectively, depended on the previous change in saccade direction. Saccade targets were most salient (Fig. \ref{FigAngle} E) and visited more by all other participants (Fig. \ref{FigAngle} D) if the previous saccade had a large change in direction (180 degree). Saliency values rose continuously with larger changes in saccade direction. Empirical density of the saccade targets is highest for return saccades (180 degree) but lowest for saccades which turn left or right (90 degree) compared to the previous saccade. Although saliency depends on the change in direction, all fixations are predicted below chance by the visual saliency model. This agrees with the notion that visual saliency does not predict fixation locations in visual search above chance \cite{henderson2007visual}. 


\subsection*{Spatial frequency spectra of fixated Locations}

Earlier analyses of eye movements during visual search reported similarities between the fixated locations and the target and it was assumed that such relationships are exploitable for the prediction of fixation locations \cite{hwang2009model}. Thus, we wanted to check whether we observe the corresponding differences between fixated and non-fixated image locations. As fixated locations, we extracted patches around the fixation locations and compared them to control patches extracted from the same locations in a different image from the stimulus set \cite{kienzle2009,judd2009learning}.

\begin{figure*}
\includegraphics[width = \textwidth]{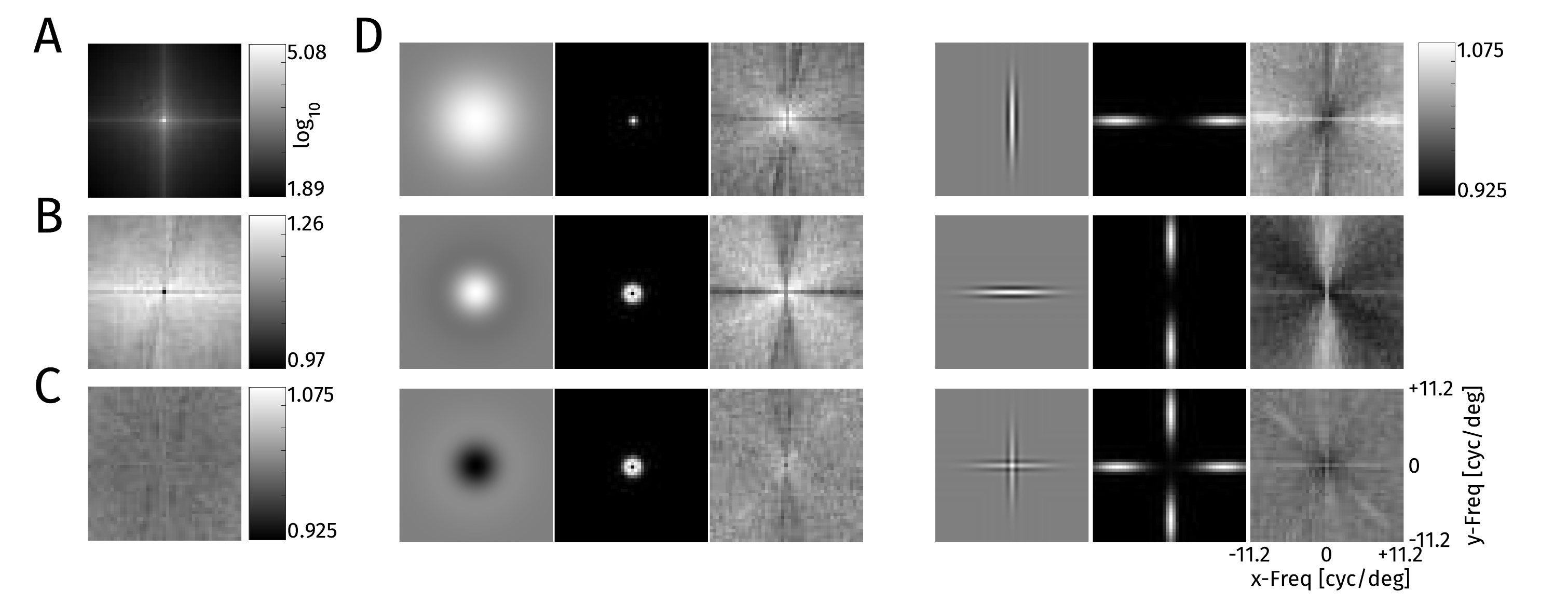}
\caption{Analysis of the spectra at fixated locations. \textbf{A}: Grand average spectrum over all fixated patches. \textbf{B}:  Spectrum from A divided by the average spectrum at control locations. The value at 0 frequency is 0.97, all other values are in the range of $[1.09,1.26]$ \textbf{C}: Average spectrum for fixations when the target is unknown, plotted as for known targets in D. \textbf{D}: Triples for each target: The target at 100\% contrast against a gray background, the Fourier space representation of the target and the average spectrum divided by the average over all targets. The color range from black to white for the third plot is always $[0.925,1.075]$. \label{fig:SearchSpectra}}
\end{figure*}

To compare the fixated patches for the different targets, we analyzed the spectra of the patches (Figure \ref{fig:SearchSpectra}). As displayed in Figure \ref{fig:SearchSpectra} A, the average spectrum of a fixated patch looks much like the spectrum of any image patch with a clear $\frac{1}{f}$ decline in spatial frequency content and a preference for horizontal and vertical structure. As these strong effects hide all other effects, all other spectra are divided by the spectra of the comparison patches for display. 

The overall spectrum of fixated patches shows increased power for all frequencies and orientations (Fig. \ref{fig:SearchSpectra} B) compared to a random image patch. This roughly means that fixated patches have more contrast than non-fixated patches.  The unknown target condition (Fig. \ref{fig:SearchSpectra} C) produces no clear deviation from the average over the conditions with known target. Searching for a specific target produces a slight bias of the fixated image patches towards being more similar to the spectrum of the target (Fig.\ref{fig:SearchSpectra} D).  The deviations of the single targets from the grand average are all smaller than $5\%$, however, while the variance over patches is substantial ($\frac{SD}{M}\in [ 78.65\%, 161.03\%]$, average $= 91.10\%$).  

While these results indicate a bias towards image patches, which have similar spectrum to the target, differences in the range of 0.1 standard deviations are certainly too small to infer the fixation category from the spectrum. Thus the only distinction, which might provide some predictive value is the generally increased contrast at fixated locations in general.


\subsection*{Target difficulty}

\begin{figure*}
\includegraphics[width = \textwidth]{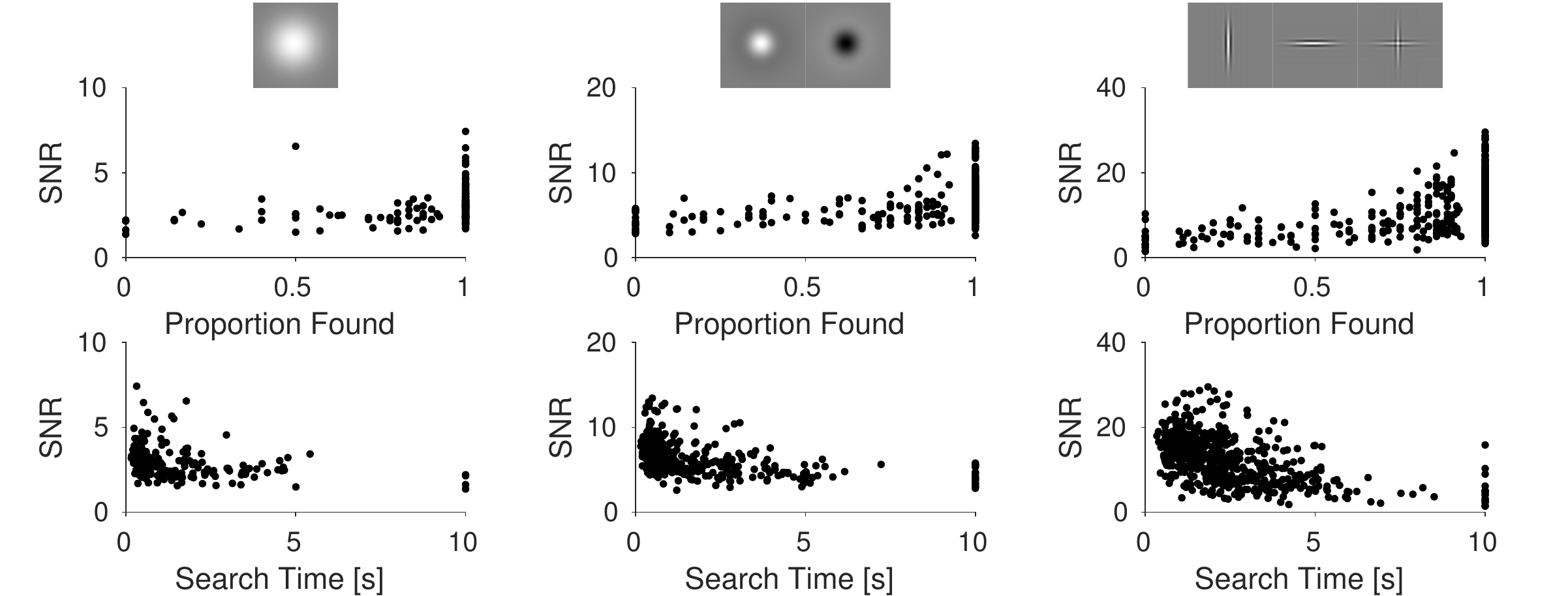}
\caption{Signal to Noise ratio from a recently published early vision model for all target-background combinations \cite{schutt2017image}. Search times and detectability are correlated, when separating analysis between target types. }{\label{fig:TargetDiff}}
\end{figure*}

Since we hid the targets on different positions in the images, it was sometimes rather easy and sometimes hard to find them. A direct measure of how difficult it is to find a target, is the time it took participants to find the target. As a computational measure, we used a recently published early vision model for images, which computes a signal-to-noise-ratio (SNR) of the target on the background for all target patches \cite{schutt2017image}. We did this to i) evaluate the models prediction to check whether the model can predict search behavior on natural scenes and ii) to have a measure of visibility for each of the targets. Figure \ref{fig:TargetDiff} shows that whether and how fast a target is found, was correlated with the predicted SNR from the early vision model. The computed SNR by the early vision thus predicts search behavior. The model computes SNRs for foveal vision. Note, that the high-spatial frequency targets have higher ratios than the low-spatial frequency targets and are thus easier to see, when looked at directly. Nonetheless, low-spatial frequency targets were found significantly faster than high-spatial frequency target, arguing that the periphery plays a highly important role in visual search \cite{cajar2016spatial}. 

\section*{Discussion} 
We studied visual search in natural scenes and found that fixation duration and saccade amplitude depend on the low-level properties of the search target. The different influences of a target on these basic eye-movement characteristics are part of a top-down search strategy, since differences between targets disappeared whenever participants did not know which target to search for. The differences in basic eye-movement characteristics can not be attributed to target difficulty or image background, because targets were searched on the same images on multiple positions. These results are important for dynamical models of eye movement control, because they show that humans adjust their basic search behavior to the target they look for. In our study, fixation durations and saccade amplitudes were longer for low-spatial frequency targets. Previous research has shown that detectability of targets in the periphery depends on spatial frequency \cite{pointer1989contrast} and fixation duration \cite{geisler1995separation}. Increasing fixation duration thus leads to a larger window of detectability and low-spatial frequency targets can generally be detected from further away. For high-spatial frequency targets, participants decreased their fixation durations strongly over time. This makes sense because i) high-spatial frequency targets have a higher SNR when looked at directly (see Figure \ref{fig:TargetDiff}) and ii) can not be detected from far away, which reduces the need for a large attentional window. 

A recently published model of early vision \cite{schutt2017image} was used to predict target difficulty. Separately looking at the results for the different targets produced promising results, because we found high correlations between a targets signal-to-noise-ratio and search times and detectability values. However, the targets themselves had very different signal-to-noise-ratios, which were not reflected in search times. This is due to the fact that the model only imitates foveal input and shows that the periphery plays a strong role when searching for artificial targets on complex backgrounds. 

Analyses of the fixation locations show that searchers slightly adjust where they look to the target, confirming earlier reports \cite{wolfe1994guided,hwang2009model}. However, the influence of the target on fixation location (investigated by comparing spatial frequency spectra) is rather small. This agrees with the notion, that participants do not merely look at positions which mostly resemble the target \cite{najemnik2005optimal}, but take their peripheral vision into account.


Post-hoc analyses of the changes in saccade direction and its dependence on further viewing behavior revealed interesting results. Saccades, which generated strong changes with respect to previous scanpath direction, landed at locations with higher empirical fixation probability and visual saliency, and were influenced by target properties. Saccades which maintained direction were not influenced by the search target and corresponding endpoints had medium fixation density and low saliency values. Regarding fixation durations, this confirms the idea that saccade generation underlies mixed control \cite{henderson2008eye,trukenbrod2014icat}, meaning that the visual input as well as some independent time-giver influences when a saccade is generated. The eyes simply progress in a default manner with {\sl saccadic momentum} saccades, unless an interesting object captures the attention, which prolongs fixation duration and urges the eyes to change direction. This idea strongly suggests, that eye movement behavior in search is shaped not only by foveal, but also peripheral input \cite{laubrock2013control,cajar2016spatial}. 

Our results suggest two different ways of searching, a selective search and a default scan that primarily moves the eyes forward in one direction. These findings agree with the study by Bays and Husain \cite{bays2012active}, who stated that return saccades are generally inhibited and only executed if a saccade target is highly interesting while forward saccades are facilitated and more frequent than a random, memoryless manner would predict. 
Eye movements play a substantial role for visual search in natural scenes and are at least partially under top-town control. However, there also seems to be a default scanning mechanism, which continues to move the eyes in the previous saccade direction and is not adjusted according to the target-template. Thus, our results are consistent with at least two mechanisms controlling eye movements under natural search conditions, which is important for dynamical models of scanpath generation \cite{engbert2015spatial,le2015saccadic,tatler2017latest}.

\section*{Methods}
As targets, we designed 6 different low level targets with different orientation and spatial frequency content (Fig. \ref{FigTargets}):

A \emph{Gaussian blob} with a standard deviation of $0.4^\circ$ of visual angle. This is an isotropic stimulus, which is a Gaussian in spatial frequency as well (with a standard deviation of $\sigma_f=0.3979$). A \emph{positive Mexican hat}, the difference between a Gaussian with a standard deviation of $0.2^\circ$ and a Gaussian with standard deviation $0.4^\circ$. This stimulus is isotropic and has a peak frequency of roughly $0.7\frac{cyc}{deg}$. A \emph{negative Mexican hat}, the negative of the positive Mexican hat, which has exactly the same spatial frequency spectrum. A \emph{vertical Gabor}, the product of $8\frac{cyc}{deg}$ vertical cosine centered at the origin and a Gaussian with standard deviations of $0.06^\circ$ and $0.32^\circ$ in x and y direction. In frequency space this stimulus is strongly oriented and has a relatively broad frequency peak at $8\frac{cyc}{deg}$. A \emph{horizontal Gabor}, the same as the vertical Gabor but oriented horizontally. A \emph{Gabor cross}, the sum of the two Gabors, each at half the contrast.

All stimuli but the Gaussian blob were near zero mean and all stimuli were normalized to have an amplitude of 1, i.e. $max(abs(T))=1$. 

To embed the targets into the natural images we first converted the image to luminance values  based on a power function fitted to the measured luminance response of the monitor. We then combined this luminance image $I_L$ with the target $T$ with a luminance amplitude $\alpha L_{max}$ fixed relative to the maximum luminance displayable on the monitor $L_{max}$ as follows:
\begin{equation}
I_{fin} = \alpha L_{max} + (1-2\alpha)I_L +\alpha L_{max} T.
\end{equation}

That is, we rescaled the image to the range $[\alpha ,(1-\alpha)]L_{max}$ and then added the target with a luminance amplitude of $\alpha L_{max}$, such that the final image $I_{fin}$ never left the displayable range. We then converted the image $I_{fin}$ back to $[0,255]$ grayscale values by inverting the fitted power function. 

\subsection*{Target Locations}
For placement of the targets we lay a grid of $4\times 2$ rectangles over each image. Within each rectangle we chose a random position for each target and image, which was at least 100 pixels away from the border, such that the target was not cut off at any side. The original plan was to present each target at each position in each image once over the eight sessions of one observer. Unfortunately, a bug in the experimental code led to a random choice of the target location instead, but we sampled only among the 8 possible locations sampled for the target-image combination. Most target-position-image combinations appeared between 6 and 10 times (10 participants and about 20 \% target absent trials, mean=7.8) and none was present more than 16 times. We are rather certain that participants could not remember the position-target-image combinations over 1200 trials and even if they did, a target appeared so rarely again at the same position that it would not have been strongly predictive of the target position. Furthermore no participant mentioned noticing anything like repeating target positions. 

\subsection*{Experiment}

\subsubsection*{Stimuli}

As stimulus material we used 25 images taken by L.R. and a member of the Potsdam lab with a Canon EOS 50 D digital camera (max. 4752 x 3168 pixels). The images were outdoor scenes without people, animals or written words present. Most images had parts with a lot of high-spatial frequency content (grass or woods) and parts with no high-spatial frequency content (sky or empty street). They were all taken on a bright sunny day in the summer.  
\subsubsection*{Stimulus Presentation}
Stimuli were presented on a 20-inch CRT monitor (Mitsubishi Diamond Pro 2070; frame rate 120~HZ, resolution 1280$\times$1024 pixels; Mitsubishi Electric Corporation, Tokyo, Japan). All pictures were reduced to a size of 1200$\times$960 pixels. For the presentation during the experiment, images were  displayed in the center of the screen with gray borders extending 32 pixels to the top/bottom and 40 pixels to the left/right of the image. The image covered 31.1 degree~of visual angle in the horizontal and 24.9 degree~in the vertical dimension.

\subsubsection*{Participants}
We recorded eye movements from 10 human participants (4 female) with normal or corrected-to-normal vision in 8 separate sessions on different days. 6 participants were students from a nearby high school (age 17 to 18) and 4 were students at the University of Potsdam (age 22 to 26). 

\subsubsection*{Procedure}
Participants were instructed to position their heads on a chin rest in front of a computer screen at a viewing distance of 70~cm. Eye movements were recorded binocularly using an desktop mounted Eyelink 1000 video-based-eyetracker (SR-Research, Osgoode/ON, Canada) with a sampling rate of 1000~Hz. Participants were instructed to search a target for the upcoming 25 images. Before each block of 25 images, the target was presented on an example image, marked by a red square. Each session consisted of 6 blocks with 25 images with the 6 different targets. The 25 images were always the same images. 

Overall 10 participants searched 6 targets on 25 images in 8 sessions, thus we collected data of 12000 search trials. Target absent trials made up between 3 and 7 for each block of 25 images ($\sim 80\%$).

Trials began with a black fixation cross presented on gray background at a random position within the image borders. After successful fixation, the image was presented with the fixation cross still present for 125~ms. This was done to assure a prolonged first fixation to reduce the central fixation tendency of the initial saccadic response \cite{tatler2007central,rothkegel2017temporal}. After removal of the fixation cross, participants were allowed to search the image for the previously defined target for 10~s. Participants were instructed to press the space bar to end the trial once a target was found. In $\sim 80\%$ of the trials the target was present. 

At the end of each session participants could earn a bonus of up to 5\euro ~additional to a fixed 10\euro~reimbursement, depending on the number of points collected. Participants earned 1 point for each correctly identified target. If participants pressed the bar although no target was present, one point was subtracted. 

\subsubsection*{Data preprocessing and saccade detection}
For saccade detection we applied a velocity-based algorithm \cite{engbert2003microsaccades,engbert2006microsaccades}. This algorithm marks an event as a saccade if it has a minimum amplitude of 0.5 degree~and exceeds the average velocity during a trial by 6 median-based standard deviations for at least 6 data samples (6 ms). The epoch between two subsequent saccades is defined as a fixation.  All fixations with a duration of less than 50~ms were removed from further analysis since these are largely glissades \cite{nystrom_adaptive_2010}. The number of fixations for further analyses was 166{,}903.

\subsection*{Fixation locations analysis}

\subsubsection*{Empirical density}
To estimate empirical fixation densities, we used kernel density estimation as implemented in the R package SpatStat (version 1.51-0). 

To estimate the bandwidth for the kernel density estimate we used leave one subject out cross-validation, i.e. for each subject we evaluated the likelihood of their data under a kernel density estimate based on the data from all other subjects repeating this procedure with bandwidths ranging from .5 to 2 degrees of visual angle (dva) in steps of 0.1 dva. We report the results with the best bandwidth chosen for each image separately.

\subsubsection*{Spatial frequency spectra}

To analyse the image properties at fixation locations we extracted image patches around fixation locations and compared them over targets and to comparison locations. We extracted $79\times 79$ pixel patches ($\approx2.05\times 2.05$ dva), around the fixated pixel, for all fixation locations for which this patch lay entirely inside the image. To obtain comparison patches, we extracted patches at the measured fixations locations shifting the image index by one. I.e. we used the fixations from picture one to extract patches from picture two etc. and the fixations from the last picture to extract patches from the first picture, as was one done earlier to train saliency models \cite{judd2009learning,kienzle2009}.
For analysis the patches were converted the patches to luminance using the measured gamma curves of the screen calculated the Fourier-Spectrum using MATLAB's fft2 function.

\bibliography{Library}

\section*{Acknowledgements}

This work was supported by Deutsche Forschungsgemeinschaft (grants EN 471/13-1 and WI 2103/4-1 to R. E. and F. A. W., resp., and CRC 1294). We thank the members of the EyeLab at the university of Potsdam for conducting the experiment and collecting the data. We thank Anke Cajar for reviewing the manuscript prior to submission and Daniel Backhaus for assisting with the creation of the stimulus material. 

\section*{Author contributions statement}

L.O.M.R and H.H.S programmed the experiment, analyzed the data and wrote the manuscript, H.A.T, F.A.F and R.E designed the experiment. All authors reviewed the manuscript.  

\section*{Competing financial interests:}  The authors declare no competing financial interests.

\end{document}